\documentclass[letterpaper]{jpconf}

\usepackage{graphicx}

\begin{document}

\title{Quantum Monte Carlo study of inhomogeneous neutron matter}
\author{Stefano Gandolfi}
\address{Theoretical Division, Los Alamos National Laboratory
Los Alamos, NM 87545, USA}
\ead{stefano@lanl.gov}

\begin{abstract}
We present an ab-initio study of neutron drops.
We use Quantum Monte Carlo techniques to calculate
the energy up to 54 neutrons in different external potentials, and we
compare the results with Skyrme forces.  We also calculate the rms radii
and radial densities, and we find that a re-adjustment of the gradient
term in Skyrme is needed in order to reproduce the properties of these
systems given by the ab-initio calculation.  By using the ab-initio
results for neutron drops for close- and open-shell configurations,
we suggest how to improve Skyrme forces when dealing with systems
with large isospin-asymmetries like neutron-rich nuclei.

\end{abstract}

\section{Introduction}

Even though neutron drops, neutrons confined by an external field, 
cannot be studied experimentally, they have nevertheless generated 
intense theoretical interests for a couple of reasons.
Neutron drops
provide a very simple model of neutron-rich nuclei, in which
the core is modeled as an external potential acting on valence
neutrons. For example in Refs.~\cite{Chang:2004,Pieper:2005,Gandolfi:2006}
neutron-rich oxygen isotopes have been modeled by neutron
drops, and in Ref.~\cite{Gandolfi:2008} the same model has
been used to study calcium isotopes.  Second, they provide a
description of inhomogeneous neutron matter that can be used as
data for calibrating model energy density functionals in several
conditions~\cite{Gandolfi:2011,Bogner:2011,Drut:2011}.

On the experimental side, new facilities plan to study the physics
of neutron-rich nuclei close to the drip line. The study of these
systems is typically done by using methods based on energy density
functionals and effective forces. These forces, like Skyrme or Gogny,
are fitted to reproduce properties of nuclear and neutron matter
combined with experimental energies of nuclei close to stability.
By constraining the density functionals to the equation of state of
nuclear and neutron matter, the symmetry energy is implicitly included
in these forces. However, the terms dependent on derivatives of proton
and neutron densities are constrained only to nuclei close to
stability, and then for very small isospin-asymmetries.

The use of these functionals to study nuclei close to the neutron drip
line requires then an important extrapolation to large isospin-asymmetries.
This extrapolation is even more drastic when the Skyrme forces are used
to study the properties of the neutron star crust, where the matter is
made by extremely neutron-rich nuclei surrounded by a sea of neutrons.
For these reasons, \emph{ab-initio} calculations of these systems
starting from accurate nuclear Hamiltonians are important to constrain
density functionals.

Important advances have been made in building very accurate Hamiltonians
describing nuclear systems. The modern nucleon-nucleon forces fit
scattering data with high accuracy~\cite{Wiringa:1995,Entem:2003}, and 
also the three-body forces are quite well constrained to reproduce
properties of light nuclei~\cite{Pieper:2001,Pieper:2008}.

Furthermore, in recent years new techniques have been developed to solve
for the ground-state of nuclear systems. In this paper
we present results obtained with Quantum Monte Carlo techniques,
i.e. the well known Green's Function Monte Carlo (GFMC) and
Auxiliary Field Diffusion Monte Carlo (AFDMC). The two different
algorithms use different variational wave functions, but they 
show similar accuracy in calculating the energy of pure neutron
systems.

\section{Nuclear Hamiltonian and Quantum Monte Carlo method}

In our model, neutrons are non-relativistic point-like particles
interacting via two- and three-body forces, that are confined by
an external potential:
\begin{eqnarray}
H = \sum_{i=1}^A\frac{p_i^2}{2m} + \sum_i V_{ext}(r_i) + 
\sum_{i<j}v_{ij}+\sum_{i<j<k}v_{ijk} \,,
\end{eqnarray}
where the external potential is a Harmonic Oscillator (HO)
\begin{equation}
V_{HO}(r)=\frac{1}{2}m\omega^2r^2 \,,
\end{equation}
or a Woods-Saxon
\begin{equation}
V_{WS}(r)=-V_0\frac{1}{1+\exp[(r -r_0)/a]} \,.
\end{equation}
In our study the parameters are $V_0 = -35.5$ MeV, $r_0=3$ fm 
and $a=1.1$ fm, and $ \hbar^2 / m = 41.44$ MeV-fm$^2$.
The parametrization of the WS has been chosen to reproduce
the properties of oxygen isotopes~\cite{Chang:2004}, and 
we consider different values of $\hbar\omega$ to change 
the density of the system.

The two body-potential that we use is the Argonne
AV8'~\cite{Wiringa:2002}, a simplified form of the Argonne
AV18~\cite{Wiringa:1995}. Although simpler to use in QMC calculations,
the AV8' provides almost the same accuracy as AV18 in fitting NN
scattering data.

In this work we use the UIX three-body force, that has been
originally proposed in combination with the Argonne AV18 and
AV8'~\cite{Pudliner:1995}. Although it slightly underbinds the energy
of light nuclei, it has been extensively used to study the equation of
state of nuclear and neutron matter~\cite{Akmal:1998,Gandolfi:2009},
and gives an equation of state of pure neutron matter stiff enough
to support a neutron star with mass higher than 2 solar 
masses~\cite{Akmal:1998,Gandolfi:2009,Gandolfi:2010,Gandolfi:2012}, as suggested by recent 
observations~\cite{Demorest:2010,Steiner:2012}.  The Illinois
three-body forces have been introduced to improve the description of
both ground- and excited-states of light nuclei with $A\le8$~\cite{Pieper:2001}, 
and produce energy spectra for $A$ up to 12 in excellent agreement with 
data~\cite{Pieper:2008}, but produce an unphysical
overbinding in pure neutron systems~\cite{Sarsa:2003}. In
neutron drops their contribution is also attractive~\cite{Carlson_inprep}.

We solve the many-body ground-state using both the Green's Function
Monte Carlo (GFMC) and Auxiliary Field Diffusion Monte Carlo 
(AFDMC)~\cite{Schmidt:1999}. The main idea of QMC methods is to evolve
a many-body wave function in imaginary-time:
\begin{equation}
\Psi(\tau)=\exp[-H\tau]\Psi_v \,,
\end{equation}
where $\Psi_v$ is a variational ansatz, and $H$ is the Hamiltonian
of the system.  In the limit of $\tau\rightarrow\infty$, $\Psi$
approaches the ground-state of $H$.  The evolution in imaginary-time
is performed by sampling configurations of the system using Monte
Carlo techniques, and expectation values are evaluated over
the sampled configurations.  For more details see for example
Refs.~\cite{Gandolfi:2009,Pudliner:1997}.

The GFMC has proven to be extremely accurate
to study properties of light nuclei. The variational wave function
includes all the possible spin/isospin states of the nucleons and
provides a good variational ansatz to start the projection in
imaginary time. The exponential growing of the spin-isospin states limits the
calculation to the $^{12}C$~\cite{Pieper:2005}.
The AFDMC method does not include all the spin/isospin states in the wave
function. The latter are instead sampled using the Hubbard-Stratonovich
transformation. Using the AFDMC the calculation can be extended up
to many neutrons, making the simulation of large neutron drops with up
to several tens of neutrons possible. In the case of homogeneous
systems the AFDMC has been used to simulate up to 114 
neutrons~\cite{Gandolfi:2009}, and up to 70 neutrons in an external well.

\section{Results}

\begin{figure}
\begin{center}
\includegraphics[width=0.7\textwidth]{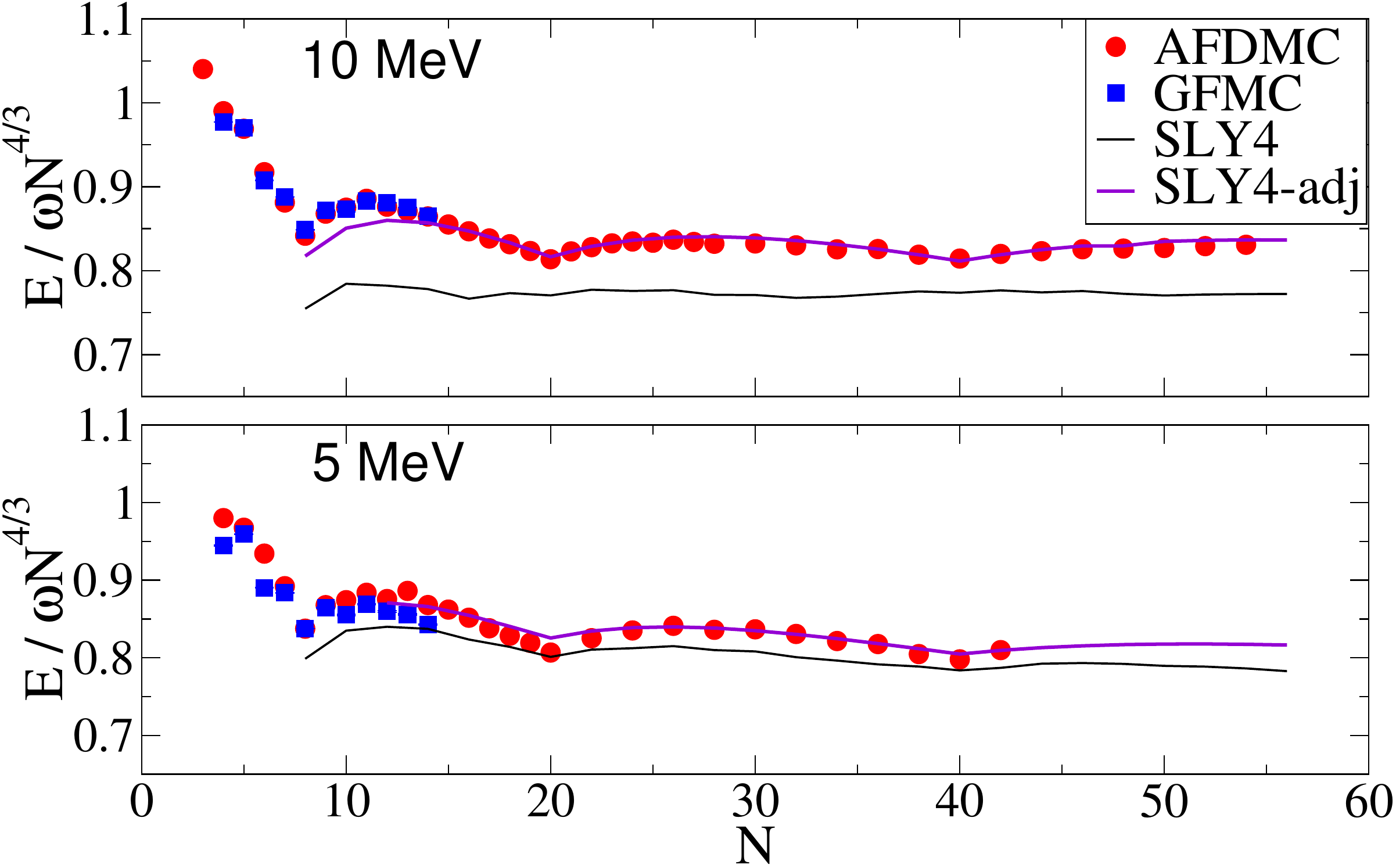}
\caption{The energy of neutrons in a HO well with $\hbar\omega=10$ MeV (upper
panel) and 5 MeV (lower panel) in units of $\omega N^{4/3}$. The red dots
are the results given by AFDMC, blue squares are from GFMC, and the black
line is the result obtained using the Skyrme SLY4. The violet line is 
the adjusted SLY4 where the strength of the gradient, pairing, and spin-orbit
terms have been changed. The figure is taken from Ref.~\cite{Gandolfi:2011}.
}
\label{fig:eho}
\end{center}
\end{figure}

The energy of neutron drops confined by $V_{HO}$ is shown in
Fig.~\ref{fig:eho} for two different frequencies of the external
potential.  The red points are the results obtained using the AFDMC
method, and the blue ones using the GFMC. The two solid lines are the
results given by using the original Skyrme SLY4 force~\cite{Chabanat:1995}, and a
modified version.  The energy is in units of the Thomas-Fermi energy,
that is proportional to $\omega N^{4/3}$, to see the extrapolation to
the thermodynamic limit.
The two QMC methods agree within 1\% for the $\hbar\omega=10$ MeV trap,
and the difference increases up to 4\% for $\hbar\omega=5$ MeV. The 
reason is that pairing correlations are included in the GFMC wave function, 
but not in the AFDMC. At low densities neutrons are
superfluids, and pairing correlations are quite important to include for 
open-shell configurations.

\begin{figure}
\begin{center}
\includegraphics[width=0.7\textwidth]{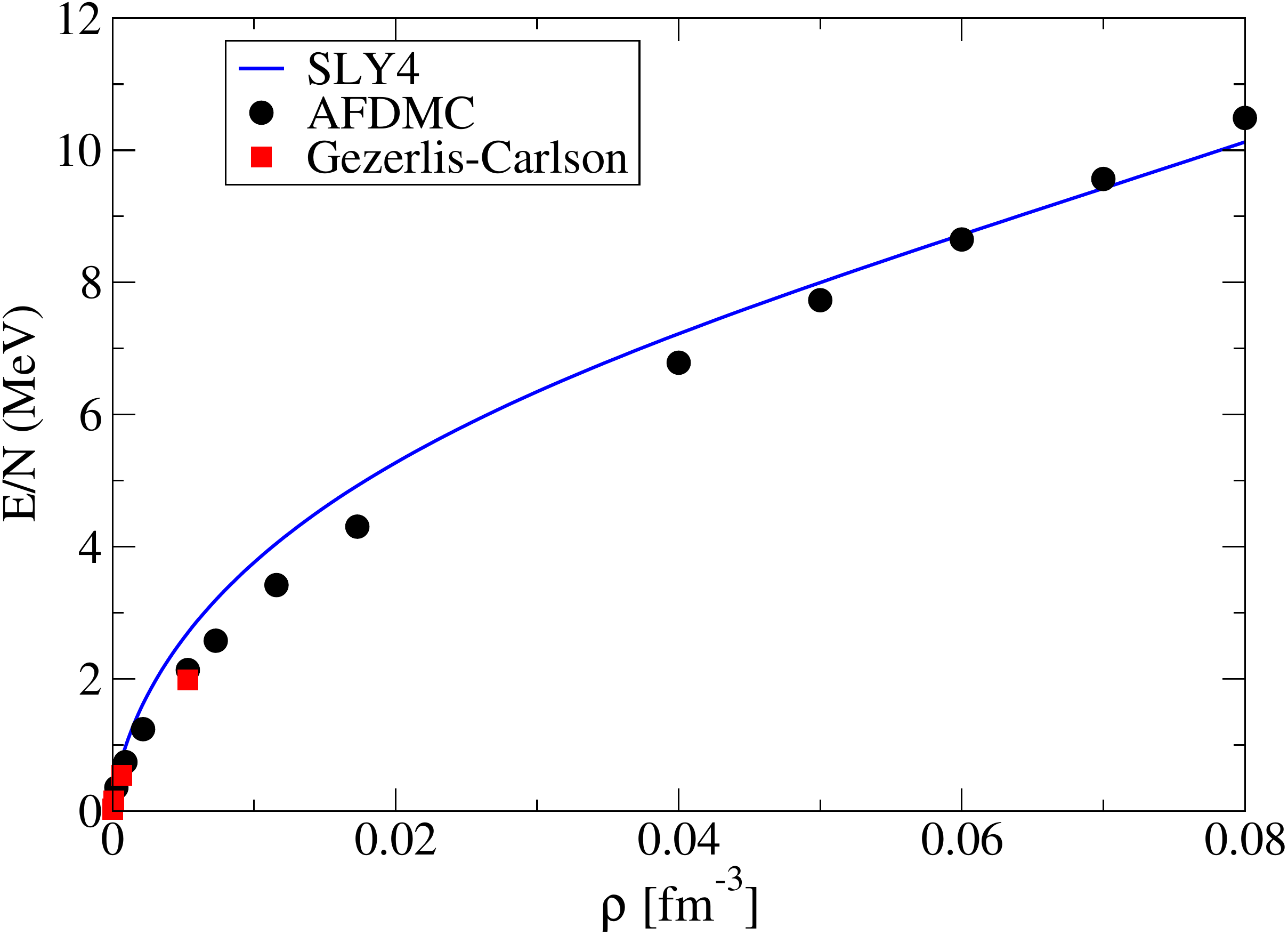}
\caption{The equation of state of pure neutron matter. The AFDMC 
results~\cite{Gandolfi:2009,Gandolfi:2008b} are compared to the 
QMC results at low densities of Gezerlis and Carlson~\cite{Gezerlis:2008,Gezerlis:2010},
and to the equation of state of Skyrme SLY4~\cite{Chabanat:1998}.
}
\label{fig:eos}
\end{center}
\end{figure}

The difference between QMC and Skyrme at closed shells is mainly due to 
two effects, the bulk contribution and the gradient term. Skyrme forces
typically give an EOS of pure neutron matter at densities lower than
saturation that is more repulsive than microscopic calculations. The equation of
state of pure neutron matter is shown in Fig.~\ref{fig:eos} where we compare
the AFDMC results from Refs.~\cite{Gandolfi:2009,Gandolfi:2008b,Gandolfi:2009b},
the GFMC calculation of Gezerlis and Carlson~\cite{Gezerlis:2008,Gezerlis:2010},
and the equation of state given by SLY4.
We make the
reasonable assumption that Skyrme's bulk term cannot explain the
difference between QMC and Skyrme energy in neutron drops.  Then, since
the pairing and the spin-orbit terms are expected to be very weak with
respect to the gradient term for closed shell configurations, we can use
the energy at N=8, 20 and 40 to re-adjust the gradient term of Skyrme.
The energy of neutron drops with N near closed shells can be used
to adjust the spin-orbit strength because for these configurations the
pairing is not important. Finally, by comparing the energy of half-filled
shells, we can tune the pairing term.

\begin{figure}
\begin{center}
\includegraphics[width=0.7\textwidth]{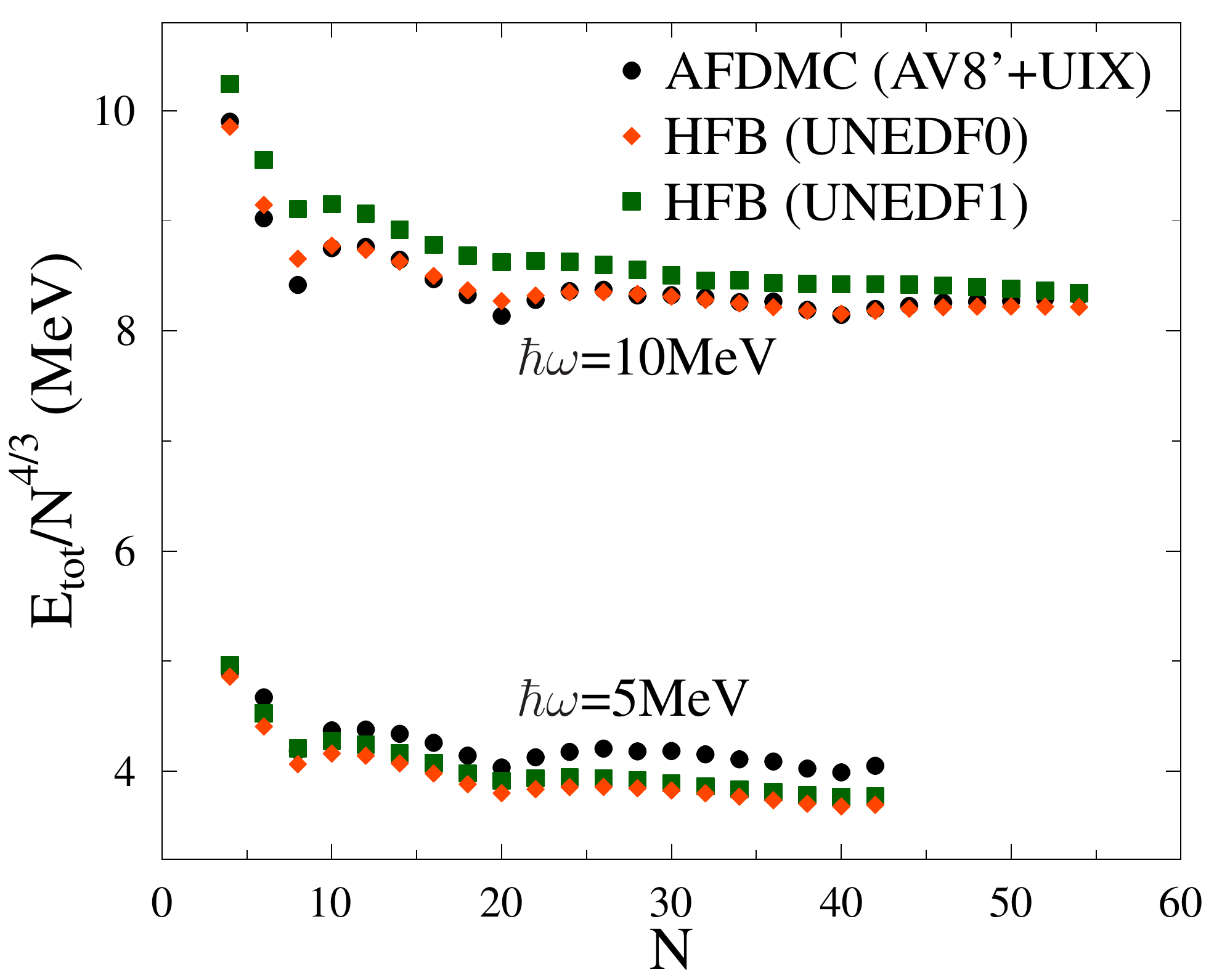}
\caption{Energy of neutrons in a HO well. The AFDMC results are compared
with those given by UNEDF0 and UNEDF1 calculations. Figure taken from 
Ref.~\cite{Kortelainen:2012} with courtesy of W. Nazarewicz.
}
\label{fig:eho_unedf}
\end{center}
\end{figure}

The ab-initio results can also be used as a benchmark
for new Skyrme forces. For example, in a recent paper Kortelainen {\emph et al.}
have derived new Skyrme forces, and they have produced neutron drops results
as a prediction to be compared with QMC calculations.
The comparison is shown in Fig.~\ref{fig:eho_unedf}.
The agreement between the two new Skyrme forces with QMC results is
remarkable considering that the ab-initio energies of neutron drops 
have not been included as a constraint in producing the UNEDF0 and UNEDF1 forces.

\begin{figure}
\begin{center}
\includegraphics[width=0.7\textwidth]{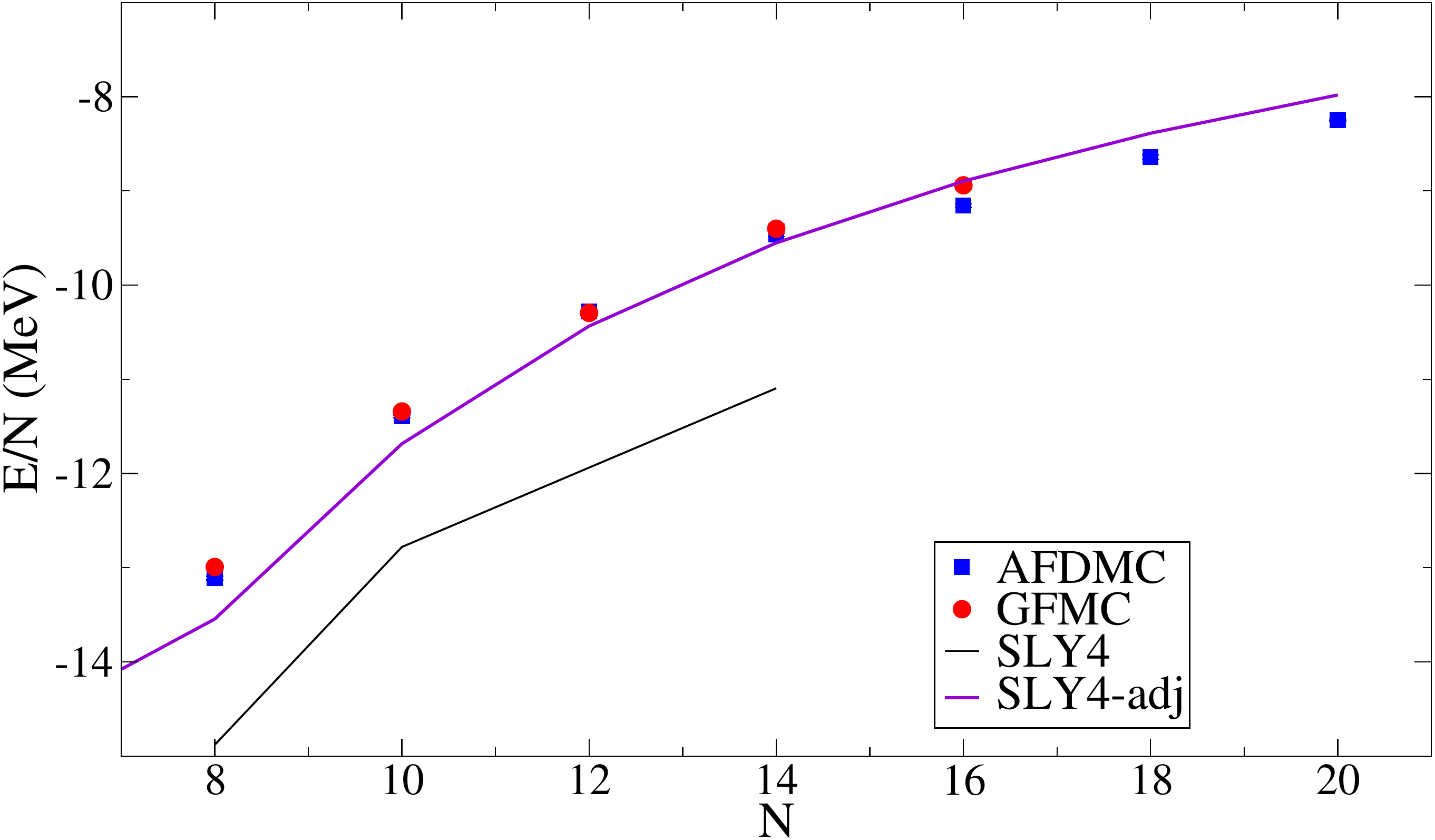}
\caption{The energy per particle of neutrons in a WS well obtained using 
AFDMC (blue squares) and GFMC (red points) compared to Skyrme.
Figure taken from Ref.~\cite{Gandolfi:2011}.
}
\label{fig:ews}
\end{center}
\end{figure}

The use of a Woods-Saxon external potential $V_{WS}$ is also interesting
because it reflects properties of nuclei. It saturates for some number
of neutrons, providing a good test of the surface effects of Skyrme. The
comparison between the original and the adjusted Skyrme SLY4 with QMC
for neutrons in a WS potential is shown in Fig.~\ref{fig:ews}. We stress
that the SLY4 has been adjusted only to reproduce the results in the
HO well, and is used here to calculate the neutron drop energies as a
prediction. Also in this case the agreement with the QMC calculations
is quite good.

\begin{figure}
\begin{center}
\includegraphics[width=0.7\textwidth]{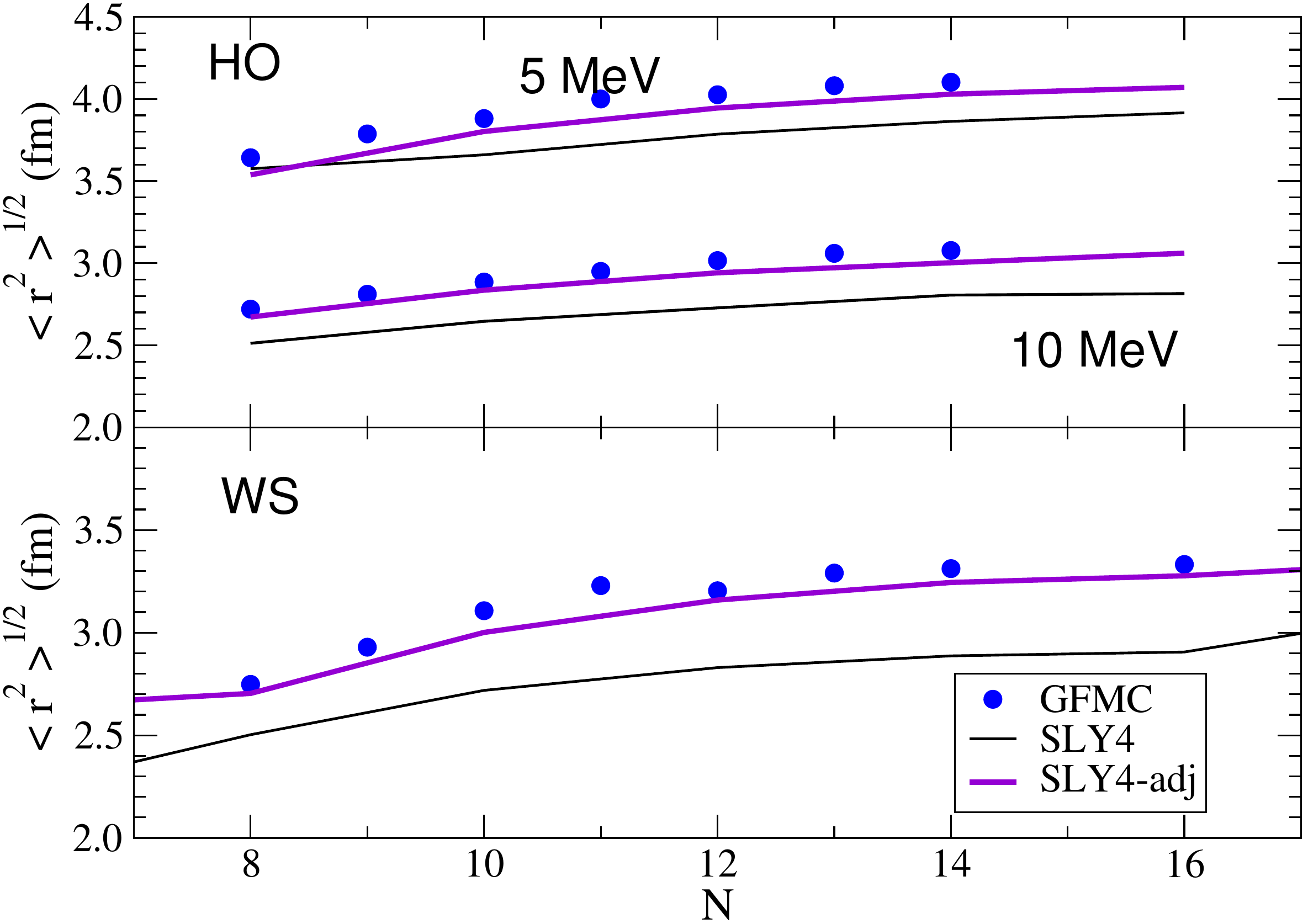}
\caption{The radii of neutron drops as a function of the number of
neutrons calculated using GFMC (blue points) compared with the original
(black line) and the adjusted SLY4 (violet line).  In the top panel
we show the results in the HO external potential, and in the bottom
panel those obtained in a WS well. Figure taken from Ref.~\cite{Gandolfi:2011}.
} 
\label{fig:radii} 
\end{center}
\end{figure}

\begin{figure}
\begin{center}
\includegraphics[width=0.7\textwidth]{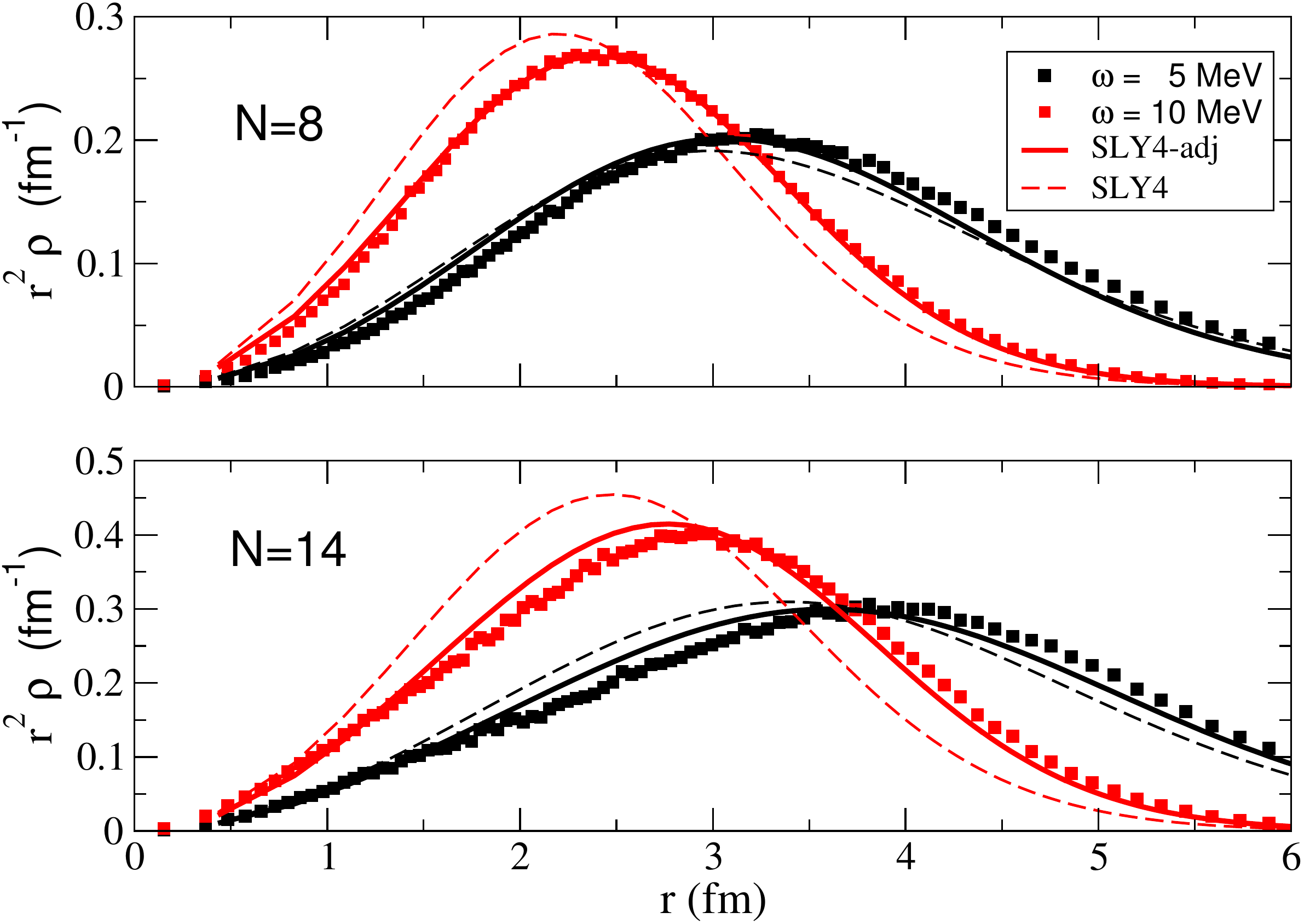}
\caption{The radial density of neutron drops for different configurations
calculated using GFMC (squares) compared with the original
(dashed lines) and the adjusted SLY4 (solid lines) in a HO well. In the top panel
we show the results of a closed shell configuration with N=8, and in the bottom
panel the density of an open shell with N=14. The two colors are for different
HO frequencies. Figure taken from Ref.~\cite{Gandolfi:2011}.
}
\label{fig:density}
\end{center}
\end{figure}

Finally, we studied the effect of the adjusted Skyrme to the structure
of neutron drops by calculating the rms radii and density. The rms
radius $\sqrt{<r^2>}$ is shown in Fig.~\ref{fig:radii} where we compare
the GFMC estimate with the original and adjusted SLY4 for different
numbers of neutrons in the WS well (lower panel), and similarly with
HO well for two different frequencies (upper panel).
In Fig.~\ref{fig:density} we show
the radial density calculated with GFMC and Skyrme in the case of HO
well for a closed shell configuration with N=8 (upper panel), and for
an open-shell with N=14 (lower panel).  Both for rms radii and for the
density the adjusted Skyrme agree much better with QMC calculations,
showing that the energy constraint is sufficient to improve
the accuracy of density functionals, whose predictive power for other
operators would drastically improve.

\section{Conclusions}

We performed ab-initio calculations of neutrons confined in external
potentials. These systems are interesting to study because they provide
a simple model to describe neutron-rich nuclei, and can be used to constrain 
energy density functionals.
By using QMC techniques, we calculated the energy of neutrons in different
potentials. We have considered both harmonic oscillator and Woods-Saxon
wells in order to provide results for systems with rather different geometries.
We paid particular attention to the comparison between the
GFMC and AFDMC methods. The agreement is quite good, at the
few percent level for the total energy of the system. 

We compared the energies of neutron drops given by QMC methods with
those given by different Skyrme forces, and found that they always
produce lower energies. Since the density of these systems is lower than
nuclear densities, we suggest that the overbinding is mainly due
to the (neutron) gradient term. We found that, in order to reproduce our
ab-initio results, the pairing and the spin-orbit terms of Skyrme need
to be changed too.

We have used AFDMC to calculate the energy up to 54 neutrons,
and GFMC to calculate the rms radii and radial densities of neutron
drops. After re-adjustment of the SLY4 parametrization, the Skyrme force
provides much better agreement for these quantities.

\ack{The author would like to thank J. Carlson and R. Schiavilla for 
critical comments on the manuscript, and W. Nazarewicz for the useful discussions 
and for the permission to show Fig.~\ref{fig:eho_unedf}.
This work is supported by DOE Grants No. DE-FC02-07ER41457
(UNEDF SciDAC) and No. DE-AC52-06NA25396, and by the LANL LDRD program.
Computer time was made available by Los Alamos Open Supercomputing, and
by the National Energy Research Scientific Computing Center (NERSC).  }

\section*{References}

\bibliographystyle{iopart-num}

\begin{thebibliography}{10}
\expandafter\ifx\csname url\endcsname\relax
  \def\url#1{{\tt #1}}\fi
\expandafter\ifx\csname urlprefix\endcsname\relax\def\urlprefix{URL }\fi
\providecommand{\eprint}[2][]{\url{#2}}

\bibitem{Chang:2004}
Chang S~Y, Morales J, Jr, Pandharipande V~R, Ravenhall D~G, Carlson J, Pieper
  S~C, Wiringa R~B and Schmidt K~E 2004 {\em Nucl. Phys. A\/} {\bf 746} 215

\bibitem{Pieper:2005}
Pieper S~C 2005 {\em Nuclear Physics A\/} {\bf 751} 516

\bibitem{Gandolfi:2006}
Gandolfi S, Pederiva F, Fantoni S and Schmidt K~E 2006 {\em Phys. Rev. C\/}
  {\bf 73} 044304

\bibitem{Gandolfi:2008}
{Gandolfi} S, {Pederiva} F and {A Beccara} S 2008 {\em European Physical
  Journal A\/} {\bf 35} 207--211

\bibitem{Gandolfi:2011}
Gandolfi S, Carlson J and Pieper S~C 2011 {\em Phys. Rev. Lett.\/} {\bf 106}
  012501

\bibitem{Bogner:2011}
Bogner S, Furnstahl R, Hergert H, Kortelainen M, Maris P {\em et~al.\/} 2011
  {\em Phys. Rev. C\/} {\bf 84} 044306

\bibitem{Drut:2011}
Drut J~E and Platter L 2011 {\em Phys. Rev. C\/} {\bf 84}(1) 014318

\bibitem{Wiringa:1995}
Wiringa R~B, Stoks V~G~J and Schiavilla R 1995 {\em Phys. Rev. C\/} {\bf 51} 38

\bibitem{Entem:2003}
Entem D~R and Machleidt R 2003 {\em Phys. Rev. C\/} {\bf 68} 041001

\bibitem{Pieper:2001}
Pieper S~C, Pandharipande V~R, Wiringa R~B and Carlson J 2001 {\em Phys. Rev.
  C\/} {\bf 64} 014001

\bibitem{Pieper:2008}
Pieper S~C 2008 {\em AIP Conf. Proc.\/} {\bf 1011} 143

\bibitem{Wiringa:2002}
Wiringa R~B and Pieper S~C 2002 {\em Phys. Rev. Lett.\/} {\bf 89} 182501

\bibitem{Pudliner:1995}
Pudliner B~S, Pandharipande V~R, Carlson J and Wiringa R~B 1995 {\em Phys. Rev.
  Lett.\/} {\bf 74} 4396

\bibitem{Akmal:1998}
Akmal A, Pandharipande V~R and Ravenhall D~G 1998 {\em Phys. Rev. C\/} {\bf 58}
  1804

\bibitem{Gandolfi:2009}
Gandolfi S, Illarionov A~Y, Schmidt K~E, Pederiva F and Fantoni S 2009 {\em
  Phys. Rev. C\/} {\bf 79} 054005

\bibitem{Gandolfi:2010}
Gandolfi S, Illarionov A~Y, Fantoni S, Miller J, Pederiva F and Schmidt K 2010
  {\em Mon. Not. R. Astron. Soc.\/} {\bf 404} L35

\bibitem{Gandolfi:2012}
Gandolfi S, Carlson J and Reddy S 2012 {\em Phys. Rev. C\/} {\bf 85} 032801

\bibitem{Demorest:2010}
Demorest P~B, Pennucci T, Ransom S~M, Roberts M~S~E and Hessels J~W~T 2010 {\em
  Nature\/} {\bf 467} 1081

\bibitem{Steiner:2012}
Steiner A~W and Gandolfi S 2012 {\em Phys. Rev. Lett.\/} {\bf 108} 081102

\bibitem{Sarsa:2003}
Sarsa A, Fantoni S, Schmidt K~E and Pederiva F 2003 {\em Phys. Rev. C\/} {\bf
  68} 024308

\bibitem{Carlson_inprep}
Carlson J, Gandolfi S, Maris P, Pieper S~C and Vary J~P  In preparation

\bibitem{Schmidt:1999}
Schmidt K~E and Fantoni S 1999 {\em Phys. Lett. B\/} {\bf 446} 99

\bibitem{Pudliner:1997}
Pudliner B~S, Pandharipande V~R, Carlson J, Pieper S~C and Wiringa R~B 1997
  {\em Phys. Rev. C\/} {\bf 56} 1720

\bibitem{Chabanat:1995}
Chabanat E, Bonche P, Haensel P, Meyer J and Schaeffer R 1995 {\em Phys.
  Scr.\/} {\bf T56} 231

\bibitem{Gandolfi:2008b}
Gandolfi S, Illarionov A~Y, Fantoni S, Pederiva F and Schmidt K~E 2008 {\em
  Phys. Rev. Lett.\/} {\bf 101}(13) 132501

\bibitem{Gezerlis:2008}
Gezerlis A and Carlson J 2008 {\em Phys. Rev. C\/} {\bf 77} 032801

\bibitem{Gezerlis:2010}
Gezerlis A and Carlson J 2010 {\em Phys. Rev. C\/} {\bf 81}(2) 025803

\bibitem{Chabanat:1998}
{Chabanat} E, {Bonche} P, {Haensel} P, {Meyer} J and {Schaeffer} R 1998 {\em
  Nuclear Physics A\/} {\bf 635} 231--256

\bibitem{Gandolfi:2009b}
Gandolfi S, Illarionov A~Y, Pederiva F, Schmidt K~E and Fantoni S 2009 {\em
  Phys. Rev. C\/} {\bf 80}(4) 045802

\bibitem{Kortelainen:2012}
Kortelainen M, McDonnell J, Nazarewicz W, Reinhard P~G, Sarich J, Schunck N,
  Stoitsov M~V and Wild S~M 2012 {\em Phys. Rev. C\/} {\bf 85}(2) 024304

\end{thebibliography}

\providecommand{\newblock}{}

\end{document}